# Quantum key distribution over 40 dB channel loss using superconducting single photon detectors

Hiroki Takesue[1], Sae Woo Nam[2], Qiang Zhang[3], Robert H. Hadfield[2], Toshimori Honjo[1], Kiyoshi Tamaki[1], and Yoshihisa Yamamoto[3]

*1. NTT Basic Research Laboratories, NTT Corporation, 3-1 Morinosato Wakamiya, Atsugi, Kanagawa, Japan.*

*2. National Institute of Standard and Technologies.*

*325 Broadway, Boulder, Colorado 80305, USA.*

*3. E. L. Ginzton Laboratory, Stanford University, 450 Via Palou, Stanford, CA94305-4088, USA.*

**Quantum key distribution (QKD) offers an unconditionally secure means of communication based on the laws of quantum mechanics [1]. Currently, a major challenge is to achieve a QKD system with a 40 dB channel loss, which is required if we are to realize global scale QKD networks using communication satellites [2]. Here we report the first QKD experiment in which secure keys were distributed over 42 dB channel loss and 200 km of optical fibre. We employed the differential phase shift quantum key distribution (DPS-QKD) protocol [3] implemented with a 10-GHz clock frequency, and superconducting single photon detectors (SSPD) based on NbN nanowire [4,5]. The SSPD offers a very low dark count rate (a few Hz) and small timing jitter (60 ps full width at half maximum). These characteristics allowed us to construct a 10-GHz clock QKD system and thus distribute secure keys over channel loss of 42 dB. In addition, we achieved a 17 kbit/s secure key rate over 105 km of optical fibre, which is two orders of**

**magnitude higher than the previous record, and a 12.1 bit/s secure key rate over 200 km of optical fibre, which is the longest terrestrial QKD yet demonstrated. The keys generated in our experiment are secure against both general collective attacks on individual photons [6] and a specific collective attack on multi-photons, known as a sequential unambiguous state discrimination (USD) attack [7].**

Since the first QKD experiment using a 1-m free space transmission line was reported over a decade ago [8], the key distribution distance in QKD experiments has continued to increase. Most of those experiments employed the Bennett and Brassard (BB84) protocol [9] with attenuated laser light as the photon source, but did not generate secure keys because of their vulnerability against a photon number splitting (PNS) attack [10,11]. In a PNS attack, an eavesdropper (Eve) undertakes a quantum non-demolition (QND) measurement on each weak coherent pulse. If Eve finds more than one photon in one pulse, she keeps one photon in her quantum memory and sends the others to Bob through her lossless transmission line. After knowing the measurement basis through the public communication between Alice and Bob, Eve can perform a projective measurement for a stored photon, and thus obtain full information about the pulse without causing any bit errors. This attack seriously limits the performance of BB84 QKD systems implemented with coherent light sources. The most obvious but technologically difficult way to prevent a PNS attack is to use a deterministic single photon source. Motivated by this reasoning, intensive research on single photon sources is being undertaken worldwide [12-14]. In fact, several BB84-QKD experiments with single photon sources have been carried out [15, 16], but the improvement in the secure key rate has been rather limited due to the residual two photon probability $P(2) \leq \frac{1}{2} g^{(2)}(0)$ of the single photon source [17]. Figure 1 shows the secure key rate vs. channel loss for BB84 systems using single photon sources with varying $g^{(2)}(0)$. The best $g^{(2)}(0)$ for an experimental single photon source is currently limited to 0.01 [12-14].

An alternative approach is to find new protocols that are robust against a PNS attack [3, 18, 19]. With the decoy-state BB84 protocol [18,19], decoy pulses are randomly inserted whose average photon number is higher than that of the signal pulses. Recently, a 107-km secure key distribution with a bit rate of about 0.1 bit/s has been reported using superconducting transition edge sensors [20].

With the DPS protocol [3], the quantum state of a single photon is defined over many pulses from a coherent laser source. Alice randomly modulates the phase of each pulse emitted from the source by $\{0, \pi\}$. The intensity of the pulse train is adjusted so that the average photon number per pulse becomes much less than one. Bob is equipped with a 1-bit delayed Mach-Zehnder interferometer, whose two output ports are followed by two single photon detectors. When the phase difference between two adjacent pulses is 0 ($\pi$), detector 0 (1) clicks. Since the average photon number per pulse is much less than one, Bob's detectors click only occasionally. Bob discloses the time instances in which he observed the clicks to Alice via public communication, while withholding which-detector information. With the time instance information and original modulation data, Alice knows which detector clicked in those time instances at Bob's site. Therefore, Alice and Bob can share an identical bit string that can be used as a key for one-time pad cryptography.

Since a QND measurement on two consecutive pulses breaks the coherence of the multiple-pulse quantum state, a standard PNS attack based on a QND measurement introduces bit errors. Eve can reduce the error probability by increasing the number of pulses simultaneously monitored by a QND measurement, but the probability of Eve obtaining the key information also decreases, because the detector click, that is a collapse of the wavefunction, occurs randomly and non-deterministically for Bob's and Eve's wavepackets. Even if Eve has the technology to undertake such a PNS attack on an arbitrary number of time slots, the fraction of information that Eve can obtain using a PNS attack is given by $2\mu$, where $\mu$ denotes the average photon number per pulse [6]. In the presence of system errors, Eve can also launch an optimal quantum measurement attack on a fraction of the photons transmitted to Bob, whose collision probability $p_{c0}$ for each bit is bounded as

$$p_{c0} \leq 1 - e^2 - \frac{(1-6e)^2}{2}, \quad (1)$$

where $e$ denotes the system's innocent bit error rate. Then, considering a two-fold attack composed of a PNS attack and an optimal quantum measurement attack, the upper bound of the collision probability of n-bit sifted key $p_c$ is given by [6]

$$p_c = p_{c0}^{n(1-2\mu)} = \left(1 - e^2 - \frac{(1-6e)^2}{2}\right)^{n(1-2\mu)}. \quad (2)$$

Thus, the compression factor $\tau$ in the privacy amplification process is calculated as

$$\tau = -\frac{\log_2 p_c}{n} = -(1-2\mu)\log_2\left[1 - e^2 - \frac{(1-6e)^2}{2}\right]. \quad (3)$$

The secure key rate $R_{secure}$ is reduced from the sifted key rate $R_{sifted}$ according to

$$R_{secure} = R_{sifted}\{\tau + f(e)h(e)\}, \quad (4)$$

where $h(e) = -e\log_2 e - (1-e)\log_2(1-e)$ is a binary entropy function, and $f(e)$ characterizes the performance of the error correction algorithm.

Figure 1 plots the secure key rates vs. channel loss for DPS-QKD using a standard coherent laser source and BB84 using single photon sources with various characteristics. Here, we assumed the same detector condition as in the experiment described below (quantum efficiency 1.4%, dark count rate 50 Hz, time window width 50 ps, and 36% reduction of effective quantum efficiency due to time window). As regards the BB84 system, we assumed an ideal implementation, namely active demodulation with no additional loss, so Bob uses only two single photon detectors. The maximum channel loss of the DPS-QKD system is larger than that of the BB84 system using a single photon source with $g^{(2)}(0) = 10^{-5}$ and an efficiency $\eta = 1$, which is far beyond the current experimental reach [12-14].

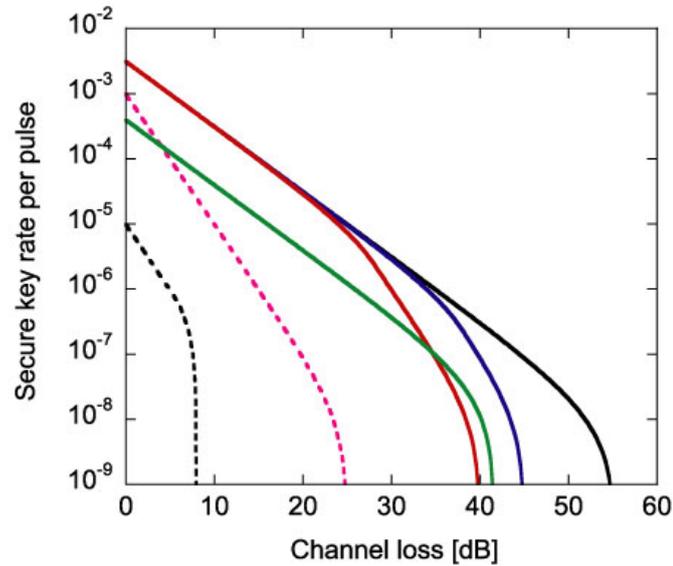

Figure 1: Secure key rate as a function of channel loss. Black solid line: BB84 with an ideal single photon source, blue line: BB84 with a single photon source with $g^{(2)}(0) = 10^{-6}$, red line: BB84 with a single photon source with $g^{(2)}(0) = 10^{-5}$, green line: DPS-QKD, pink line: BB84 with a single photon source with $g^{(2)}(0) = 10^{-2}$, black dotted line: BB84 with an attenuated laser source.

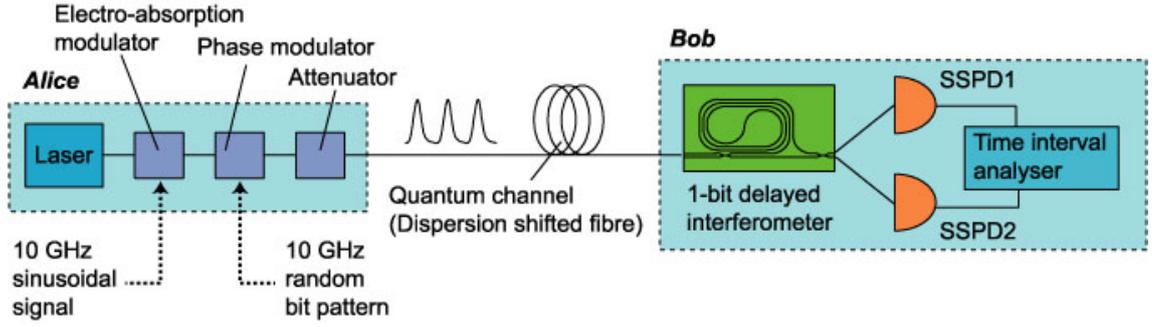

Figure 2: Experimental set-up for 10-GHz clock DPS-QKD.

Figure 2 shows the system configuration of the 10-GHz clock DPS-QKD system. A continuous wave output from a laser is transformed into a 10-GHz clock pulse train by an InGaAsP electro-absorption (EA) modulator. We generated pulses with a full width at half maximum of 15 ps. The phase of each pulse is modulated by a phase modulator driven by a 10-GHz pseudo random bit pattern from a high-speed pulse pattern generator. The average photon number per pulse is adjusted to 0.2 by an optical attenuator. The quantum channel is a dispersion shifted fibre (DSF) or a single attenuator. Bob is equipped with a 1-bit delayed interferometer fabricated using planar lightwave circuit technology. The excess loss of the interferometer is 2.5 dB. Each output port of the interferometer is connected to an SSPD. The photon detection time instances and which-detector information are recorded using a time interval analyser (TIA). In our experiment, sifted keys were actually generated between Alice and Bob, and the error rate was measured by directly comparing Alice's key with Bob's key. For each data point described below, we undertook five runs of QKD sessions, and the error rates and sifted key rates are the average of the five runs. The secure key rates were calculated by plugging the experimentally obtained error rates and sifted key rates into Eq. (4).

Figure 3 (a) shows a close-up image of an SSPD, which consists of a 100-nm wide, 4-nm thick NbN superconducting wire. The SSPD is coupled to a 9-μm core single-mode fibre as shown in Fig. 3 (b). The packaged detector is housed in a closed-cycle cryogen-free refrigerator with an operating temperature of 3 K, for convenient use in quantum information experiments [21]. The detector operates as follows: the superconducting wire is current-biased slightly below its critical current. When a photon hits the wire, a resistive hot spot is formed. Then the current density around the spot increases and eventually exceeds the critical current. As a result, a non-superconducting barrier is formed across the entire width of the wire, and a voltage pulse is formed. By discriminating the starting edge of the voltage pulses, we can measure the photon arrival time with a high timing resolution. The quantum efficiency and dark count rate of the SSPD vary when the bias current is changed. The single photon counting mechanism of an avalanche photodiode (APD) is complex (consisting of absorption, diffusion and avalanche), which results in excess dark counts and non-Gaussian timing jitter characteristics. On the other hand, the principle of SSPD is rather simple as explained above, which makes the dark count rate and timing resolution characteristics of the SSPD superior to those of an APD. We measured the timing jitter of the SSPD by launching 10-ps pulses. Figure 3(c) compares the obtained histogram of the photon arrival time with that of a single photon detector based on a frequency up-converter followed by a Si APD, which was used in our previous DPS-QKD experiments [22]. Here, the blue squares and the line denote the histogram for the SSPD, and the red line is that for the up-conversion detector. Even though the full width at half maximum (FWHM) of the jitter was approximately 60 ps, which is larger than that of the up-conversion single photon detectors, the histogram fits very well with Gaussian, and does not have a long tail, as observed in similar measurement for the up-conversion detectors [22]. Therefore, we can reduce the error probability caused by inter-symbol interference by using SSPD. In addition, the dark count rate of the SSPD was measured to be <10 Hz (typically a few Hz) when the quantum efficiency was set at 0.7%, which is much smaller than that of the up-conversion detector operated in a similar quantum efficiency (350 Hz at 0.4% quantum efficiency [22]).

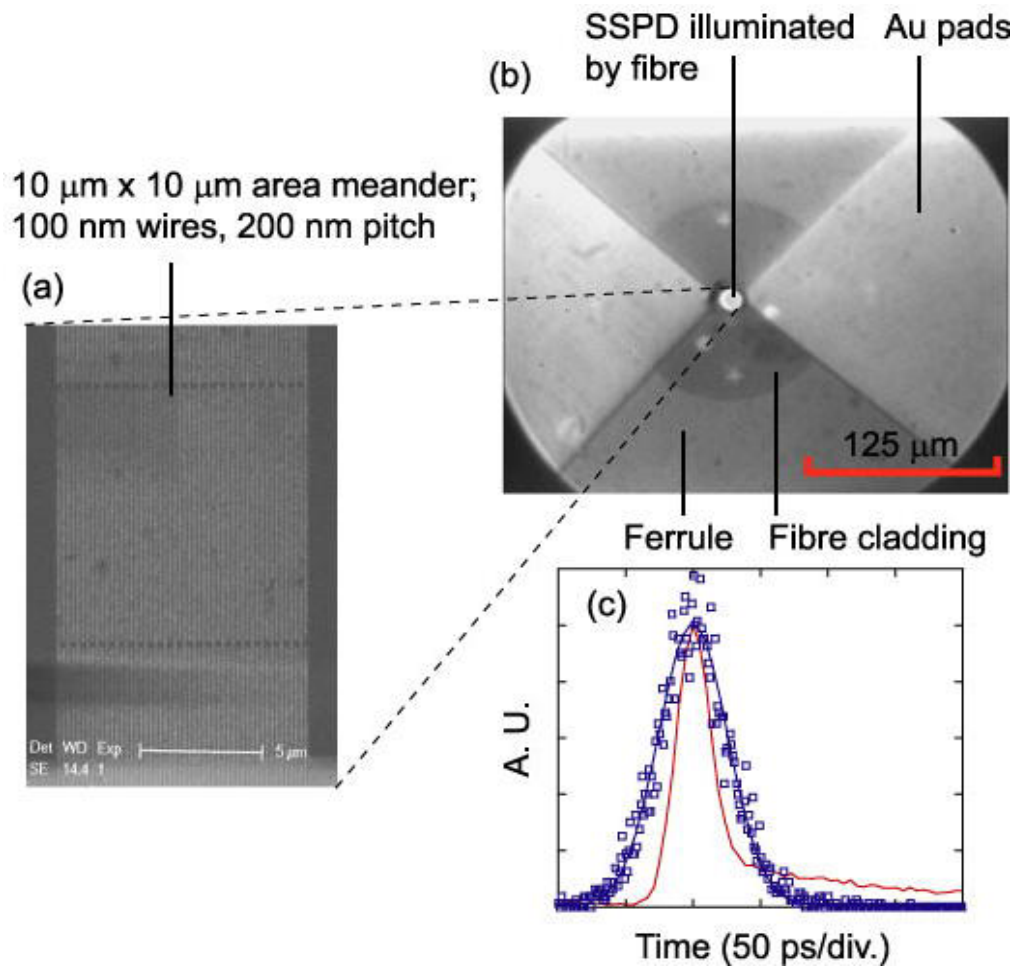

Figure 3: (a) SSPD close-up image observed with scanning electron microscope (b) Fibre alignment under optical microscope (c) Histograms of photon arrival time of SSPD (squares and blue line) and up-conversion detector (red line) when 10-ps pulse is input.

We further apply a narrow time window to the obtained time-instance data to reduce the contributions of the dark counts and the inter-symbol interference caused by neighbouring signals. To demonstrate the effectiveness of this technique, we obtained a histogram of the photon arrival time detected by one of two SSPDs. Here, the quantum efficiency and dark count rate of the SSPD were set at 0.7% and <10 Hz, respectively, and the channel loss was 31.7 dB. Figure 4 (a) shows the histogram when no time window was used. We observed a signal pulse overlap caused by the detector timing jitter. However, the peaks were well separated when we employed a 10-ps time window (Fig. 4 (b)).

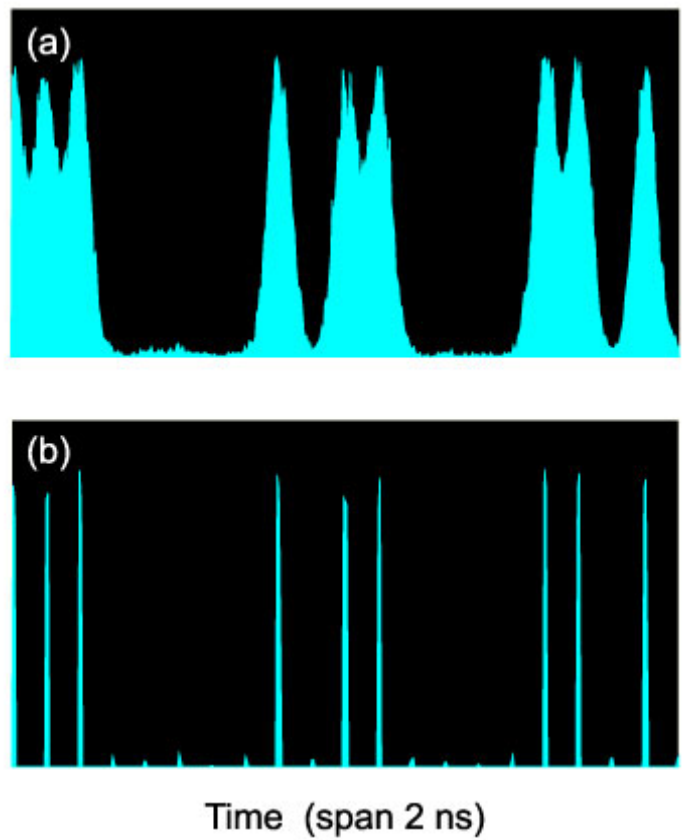

Figure 4: Histogram of the received 10-GHz clock signal. (a) without time window, (b) with 10-ps time window.

In the QKD experiment, we set the quantum efficiency, dark count rate and time window width at 1.4%, 50 Hz and 50 ps, respectively. The use of the time window reduced the effective quantum efficiency by 36%. The obtained secure key rates are shown by squares in Fig. 5, where filled and open symbols show fibre transmission points and simulated points with an optical attenuator, respectively. The solid line shows the theoretical curve calculated assuming a quantum efficiency, dark count rate and baseline system error of 1.4%, 50 Hz and 2.3%, respectively. We also undertook a DPS-QKD experiment with a 1-GHz clock, and the result is shown by the triangles in Fig. 5 [23]. The present result significantly outperforms previous QKD experiments both in secure key generation rate and distribution distance. At 105 km, we successfully generated secure keys at a rate of 17 kbit/s, which is two orders of magnitude greater than the previous record (166 bit/s at 100 km) [22]. In a 200 km fibre transmission experiment, we were able to generate secure keys with a bit rate of 12 bit/s. The maximum channel loss for secure key generation was 42.1 dB, which is more than 20 dB greater than that of previous long-distance QKD experiments.

Thus far, we have discussed security based on general collective attacks for individual photons. Now, we consider security against a sequential USD attack, which was proposed in [7]. The basic idea behind this type of attack is summarized here. First, we consider the worst case by assuming that Eve has a local oscillator that is phase-locked to the coherent light source owned by Alice. Thanks to this local oscillator, with a probability of $1 - \exp(-2\mu)$ Eve can unambiguously discriminate whether each pulse is phase-modulated by 0 or $\tilde{\pi}$ When Eve obtains $m$ $(>M)$ successful sequential measurement outcomes, she constructs $m$ trains of coherent pulses with phase modulations that depend on the measurement outcomes, and resends them to Bob. If Eve obtains $m$ $(=M)$ successful sequential measurement outcomes, with probability $p$ she resends m trains of coherent pulses with the corresponding phase modulations, and with a probability of 1-$p$ she resends a vacuum. Finally, if Eve obtains $m$ $(<M)$ successful sequential measurement outcomes, she resends a vacuum. In this attack, no error occurs inside the pulse train, but the boundary of pulse and vacuum causes a random error. Note that a high channel loss gives Eve an advantage because the number of consecutive successful USD attacks is not necessarily large. Thus, this sequential USD attack can be a potential threat when the channel loss is large.

To investigate the security against sequential USD attack, we also calculated the error threshold for this attack. The error threshold for a 200 km transmission using SSPDs with 1.4% quantum efficiency (with 36% reduction of effective quantum efficiency by the time window) was 4.74%. With μ=0.2, the error threshold for generating keys that are secure against general collective attack for individual photons is approximately 4.1%. This means that for our DPS-QKD experiments with SSPDs, general collective attacks on individual photons [6] gives a tighter security bound than a sequential USD attack [7]. Therefore, the experimental data shown by squares in Fig. 5 were all secure against both general collective attacks on individual photons and a sequential USD attack.

In conclusion, we described a 10-GHz clock differential phase shift quantum key distribution experiment using SSPDs. The small dark count rate and small timing jitter of the SSPD enabled us to generate secure keys with a 10-GHz clock system for the first time. We have distributed secure keys against general collective attacks for individual photons and a sequential USD attack over 200 km of fibre and with a 42.1 dB channel loss. In addition, we achieved a 17 kbit/s secure key rate at 105 km, which is two orders of magnitude larger than the previous record. We expect that this result to constitute an important step towards realizing both inter-city terrestrial QKD systems and global QKD systems using communication satellites that normally require a 30-35 dB channel loss [2,24].

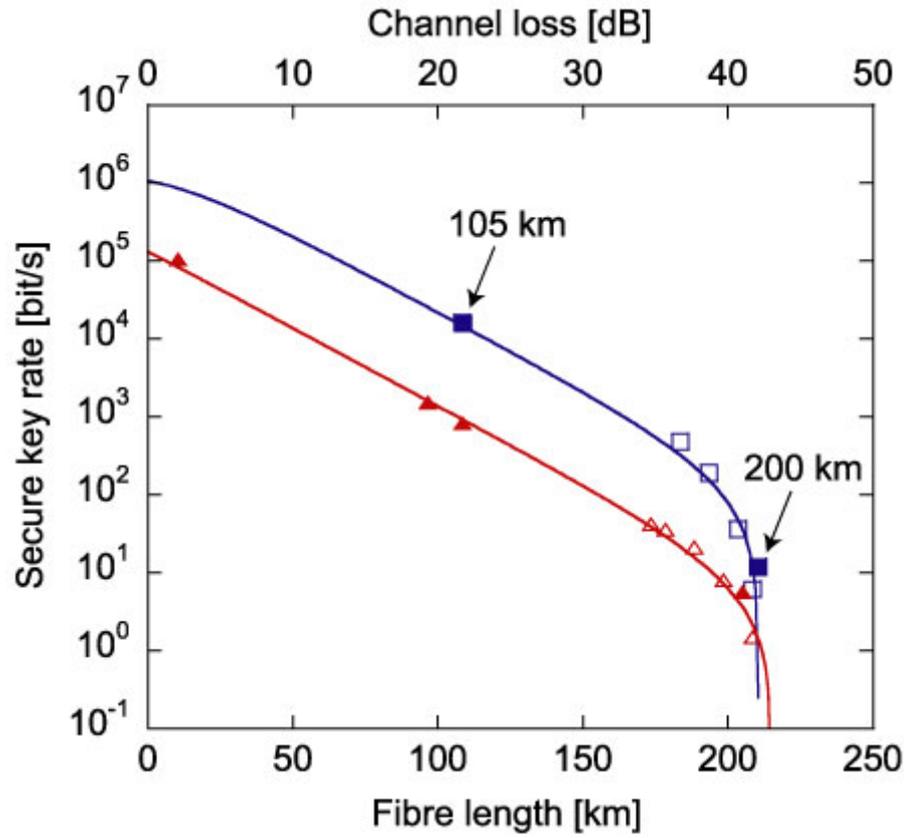

Figure 5: secure key rate as a function of fibre length with 0.2 dB/km loss. The squares and triangles show secure key rates measured generated by 10-GHz and 1-GHz clock systems with SSPDs. The filled and open symbols denote fibre transmissions and optical attenuation, respectively.

The authors thank E. Diamanti, M. M. Fejer, G. N. Gol'tsman, E. Ip, J. M. Kahn, G. Kalogerakis, L. G. Kazovsky, N. Y. Kim, C. Langrock, R. V. Roussev, and Y. Tokura for their support during this research. Financial support was provided by the CREST and SORST programs of the Japan Science and Technology Agency (JST), the National Institute of Information and Communications Technology (NICT) of Japan, the MURI Center for Photonic Quantum Information Systems (ARO/ARDA DAAD19-03-1-0199), DTO, DARPA, and the NIST Quantum Information Science Initiative.

Correspondence and requests for materials should be addressed to Hiroki Takesue (e-mail: htakesue@will.brl.ntt.co.jp).